\documentstyle[12pt,epsfig,wrapfig,rotating]{article} 
\begin{document}
 
\begin{titlepage}
\medskip
 
\vspace*{2cm}
 
\medskip
\center{\Large\bf Relativistic Heavy-Ion Physics: Experimental Overview}
\vskip 2cm
\normalsize
 
\begin{center}
Itzhak Tserruya  \footnote{e-mail: Itzhak.Tserruya@weizmann.ac.il.\\ 
                         Work supported by the Israeli Science Foundation and 
                         the Nella and Leon Benoziyo Center of High Energy Physics Research.}\\ [0.5cm]

Department of Particle Physics, Weizmann Institute,\\  
Rehovot 76100, Israel \\ [0.5cm]
 	
\end{center}
 
\vspace{1.0cm}

\begin{abstract} 
 The field of relativistic heavy-ion physics is reviewed with emphasis on new results and
highlights from the first run of the Relativistic Heavy-Ion Collider at BNL and 
the 15 year research programme at the SPS at CERN  and the AGS at BNL.
\end{abstract}
 
\vspace{3.5cm}
\normalsize
\noindent
Invited Talk at the Fourth International Conference on Physics and Astrophysics of
Quark Gluon Plasma, (ICPAQGP-2001) Jaipur, India, November 26-30, 2001.
 
\end{titlepage}


\section{INTRODUCTION}

The field of relativistic heavy-ion collisions is at a unique time in its development. 
  The Relativistic Heavy-Ion Collider (RHIC) at BNL started regular operation in the
summer of year 2000. In a very short, but extremely successful, run with an integrated 
luminosity of only a few $\mu b^{-1}$, the four RHIC experiments delivered an 
impressive amount of results offering a first glimpse at the physics of Au-Au
collisions at the energy of $\sqrt{s_{NN}} = 130~GeV$, almost one order of magnitude 
larger than the highest energy of $\sqrt{s_{NN}} = 17.2~GeV$ available at CERN. Some of these
results are new and exciting, some are puzzling and some follow the pattern already established
at lower energies. A lot more is expected to come from the second run at the design energy 
of $\sqrt{s_{NN}} = 200~GeV$, completed by the time of this conference with 
almost two orders of magnitude larger integrated luminosity.  
In addition to that, the 15 year programme at CERN SPS and BNL AGS has produced a 
wealth of very interesting and intriguing results and more is still to be expected from the SPS 
in the next few years. This combination results in very exciting opportunities for the 
study of matter under extreme conditions and in particular for the quest of the phase 
transition associated with quark-gluon plasma formation and chiral symmetry restoration. 
 
   About a dozen of different experiments at the AGS and SPS and four experiments at RHIC are or 
have been involved in this endeavour covering a very broad range of observables.
 In the limited space  of this paper it is not possible to do justice to the vast amount of
available results \cite{qm01}. A selection is therefore unavoidable and this review 
is restricted to selected topics on recent results and highlights of the field on:
(i) global observables where systematic comparisons are made from AGS up to RHIC 
energies (Section 2), (ii) hadron spectra and yields (Section 3), (iii) excess emission of 
low-mass lepton pairs (Section 4) and $J/\psi$ suppression (Section 5)
which are among the most notable results hinting at new physics from the 
SPS programme (results on these two topics are not yet available at RHIC as they require 
a much higher luminosity than achieved in the first year), and (iv) suppression of large
$p_T$ hadrons (Section 6) which is one of the highlights of the first RHIC run pointing to 
new phenomena opening up at RHIC energies.

\section{GLOBAL OBSERVABLES}
\vspace*{-0.3cm}
\subsection{\underline {$N_{ch}$ and $E_T$ distributions}}
   Global observables, like multiplicity and transverse energy, 
provide very valuable information. Besides defining the collision geometry, they 
can be related to the initial energy density, e.g. using the well-known Bjorken 
relation \cite{bjorken}, shed light into the mechanisms of particle production\cite{wang-miklos}
and provide constraints to the many models aiming at describing these collisions \cite{bass-qm99}.  

   The charged particle rapidity density at mid-rapidity has been measured by the four 
RHIC experiments -BRAHMS, PHENIX, PHOBOS and STAR- with very good agreement among 
their results \cite{phobos-200}. For central Au-Au collisions at $\sqrt{s_{NN}} = 130~GeV$, the global 
average is $dN_{ch}/d\eta = 580 \pm 18$.
 For the transverse energy rapidity density there is
only one measurement, by PHENIX, with a value $dE_T/d\eta = 578 ^{+26}_{-39}~GeV$  for the most central
2\% of the inelastic cross section \cite{phenix-ppg002}.
Using the Bjorken formula \cite{bjorken} this 
translates into an initial energy density of $\epsilon = 5.0~GeV/fm^3$ \footnote{using the 
canonical formation time $\tau = 1~fm/c$. Much shorter formation times, as advocated in saturation models
\cite{kharzeev-nardi-0012025}, would result in energy densities as high as  
$\epsilon \simeq 20~GeV/fm^3$.} i.e. 60-70\% larger than the corresponding values at the 
full SPS energy.

\begin{wrapfigure}[16]{r}{5.0cm}
\epsfxsize=5.0cm
\vspace*{-0.5cm}
\epsfbox{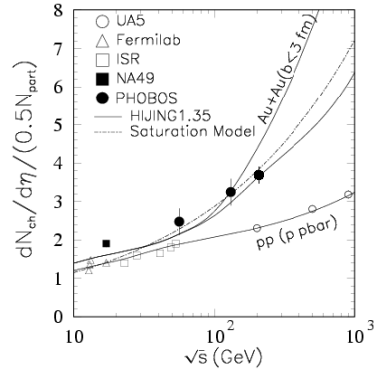}   
\caption {Energy dependence of
           charged particle density in central 
           $AA$ and $pp$ collisions~\protect\cite{wang-miklos}.}  
\end{wrapfigure} 
The energy dependence of the charged particle density $dN_{ch}/d\eta$, normalized to the number
of participating nucleon pairs ($N_{part}/2$), exhibits an almost logarithmic rise with  $\sqrt{s_{NN}}$  
from AGS up to RHIC energies, as shown in Fig.~1. However, the dependence is very different from 
the $p{\overline p}$ systematics also shown in Fig.~1. At $\sqrt{s_{NN}} = 200~GeV$, $\sim$65\%  
more particles per pair of participants are produced in central Au-Au collisions than in $p{\overline p}$. 
 Note also that the large increase predicted by the HIJING model 
with jet quenching \cite{wang-miklos} (upper curve in Fig. 1) is not observed, and from 
$\sqrt{s_{NN}} = 130$ to $200~GeV$ the particle density increases by only 15\%  \cite{phobos-200}.      
(See also Sections 2.2 and 6 for further discussion on jet quenching).
  
 The centrality dependence of the charged particle density has been proposed as a sensitive
tool to shed light on the particle production mechanism: soft processes  are believed 
to scale with the number of participants $N_{part}$, whereas hard processes, expected to 
become more significant as the energy increases, lead to a scaling with the number of binary 
collisions $N_{bin}$. PHENIX  \cite{phenix-ppg001} and PHOBOS \cite{phobos-0201005} have reported 
an increase of $dN_{ch}/d\eta$ and $dE_T/d\eta$
strongly than linear with  $N_{part}$ (the PHENIX results are shown in Fig. 2 left panel) and in the 
framework of models with these two components, like 
HIJING \cite{kharzeev-nardi-0012025,hijing}, such an increase is interpreted as evidence of 
the contribution of hard processes to particle production \footnote {This is, however, not a 
unique interpretation. The centrality dependence results of PHENIX shown in Fig. 2 have also 
been explained with a model based on purely soft processes \cite{capella-sousa-0101023}.}. This 
is to be contrasted to SPS results where WA98 reports a much weaker increase 
(also shown in Fig. 2) \cite{wa98-cent} and WA97 an even weaker increase consistent within errors 
with proportionality with  $N_{part}$ \cite{wa97-cent}. 
Both the PHENIX and WA98 data, extrapolated to peripheral collisions, are in very 
good agreement with the $p{\overline p}$ result at the same $\sqrt{s}$ derived from the UA5
data \cite{ua5}. The increased role of hard processes at higher energies
and their scaling with $N_{bin}$ could then explain the stronger increase of particle 
production in $AA$ collisions compared to  $p{\overline p}$ previously discussed in the context 
of Fig. 1. The importance of hard processes at RHIC energies is a very interesting issue which 
will be further discussed in this review.

   More surprising is the behavior of the ratio  ($dE_T/d\eta$) / ($dN_{ch}/d\eta$), the average transverse 
energy per charged particle, shown in Fig. 2 (right panel). 
This ratio is found to be independent of centrality  and approximately equal to $\sim0.8~GeV$.
Within errors of the order of 10-20\%  this ratio is also independent 
of $\sqrt{s_{NN}}$ from AGS up to RHIC, implying a constant or a very moderate increase of the 
average $p_T$ per particle.

\begin{figure}[h!]
\vspace{-0.3cm}
\begin{minipage}[h!]{55mm}
\hspace*{1.0cm}
\epsfig{file=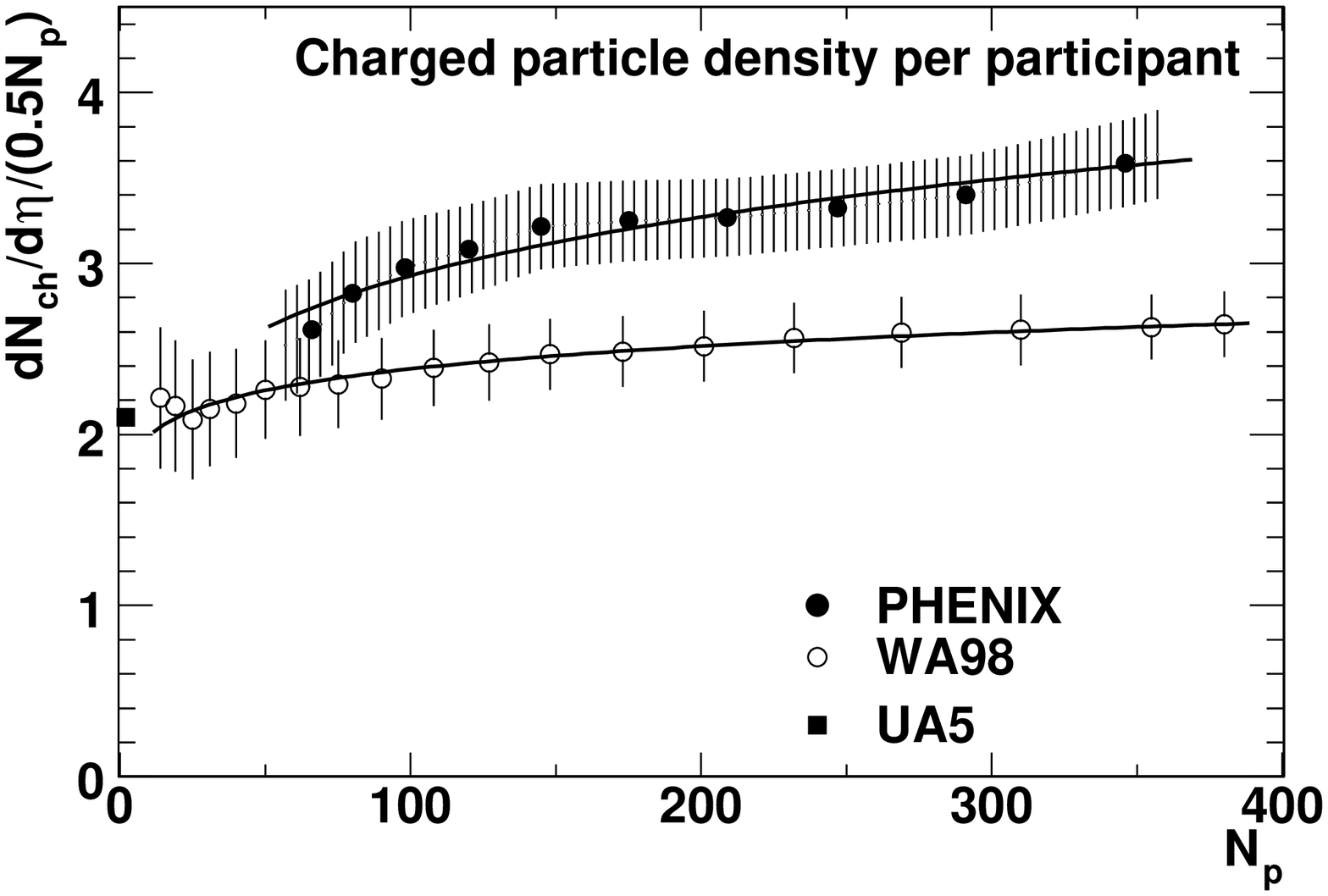,width=5.5cm,height=5.0cm}
\end{minipage}
\begin{minipage}[h!]{55mm}
\hspace*{3.0cm}
\epsfig{file=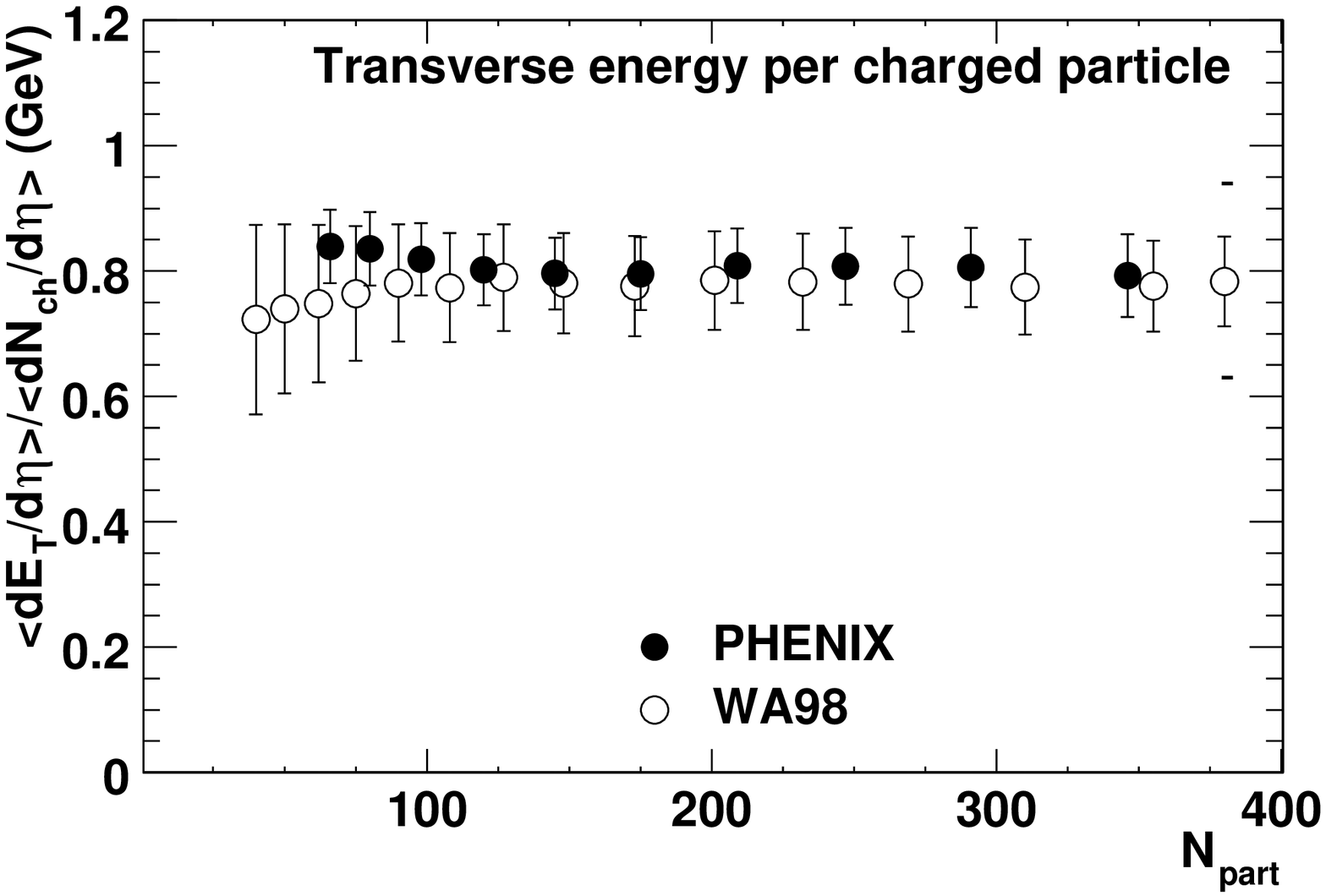,height=5.0cm,width=5.5cm}
\end{minipage}
\caption {Left panel: Centrality dependence of the charged particle density from 
          PHENIX and WA98 ~\protect\cite{phenix-qm01-milov}. Right panel: Average $E_T$ per 
          charged particle from PHENIX and WA98 ~\protect\cite{phenix-ppg002}.}
\vspace{-0.5cm}
\end{figure}

   It is also interesting to note that UA1 quotes a very similar ratio 
at $\sqrt{s} = 200~GeV$ \cite{ua1} but with a rather large error of $\sim 20\%$. This seemingly 
universal behaviour of constant energy/particle 
is one of the most puzzling results. The increase in $dE_T/d\eta$ with  $\sqrt{s_{NN}}$  translates 
into an increase in the number of produced particles rather than in the production of particles 
with larger $E_T$. 

\vspace*{-0.3cm} 
\subsection{\underline {Elliptic flow}}
     Elliptic flow originates from the spatial anisotropy
of the diamond shaped system formed in non-central collisions, later converted into 
momentum anisotropy  if enough rescatterings among particles
occur in the further evolution of the system. Elliptic flow provides therefore an opportunity 
to learn about the degree of thermal equilibrium achieved with emphasis on the early collision dynamics.
 
   Elliptic flow has been measured at RHIC by STAR \cite{flow-star}, PHOBOS \cite{flow-phobos} and 
PHENIX  \cite{phenix-ppg004} using different analysis methods and with good agreement between their results. 
The magnitude of elliptic flow, quantified e.g. by the second Fourier coefficient $v_2$ of 
the azimuthal particle distribution with respect to the reaction plane, is shown in Fig. 3 (left panel) 
as function of beam energy. The strength of $v_2$ at RHIC follows the systematic trend established over a
very broad range of lower energies from SIS to SPS.  The STAR value of 6\% is approximately 50\% 
larger than at the SPS, suggesting a much stronger early approach to local thermal equilibrium at RHIC. 
 
\begin{figure}[h!]
\vspace{-0.3cm}
\begin{minipage}[h!]{55mm}
\hspace*{1.0cm}
\epsfig{file=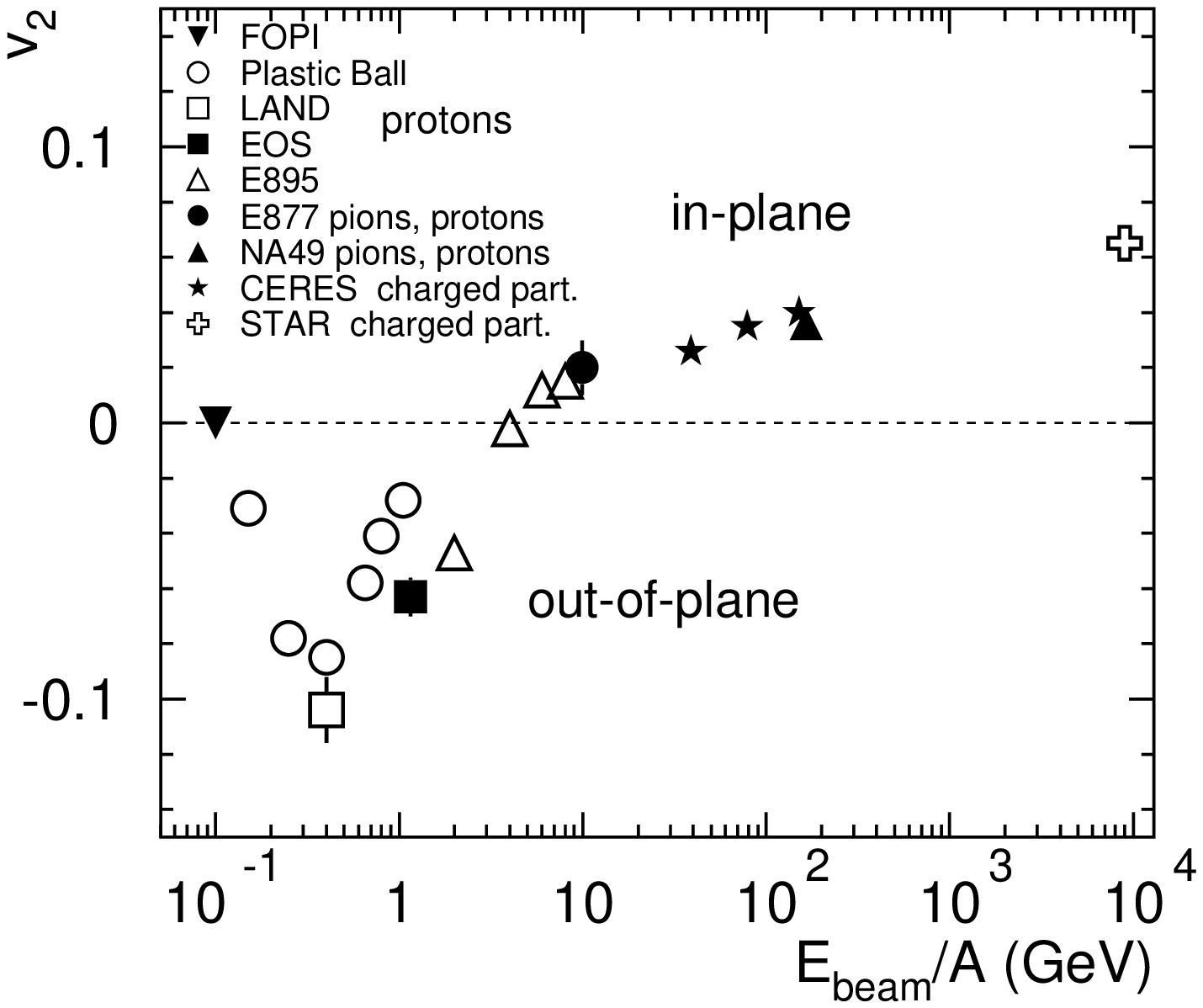,width=5.5cm,height=5.0cm}
\end{minipage}
\begin{minipage}[h!]{55mm}
\hspace*{3.0cm}
\epsfig{file=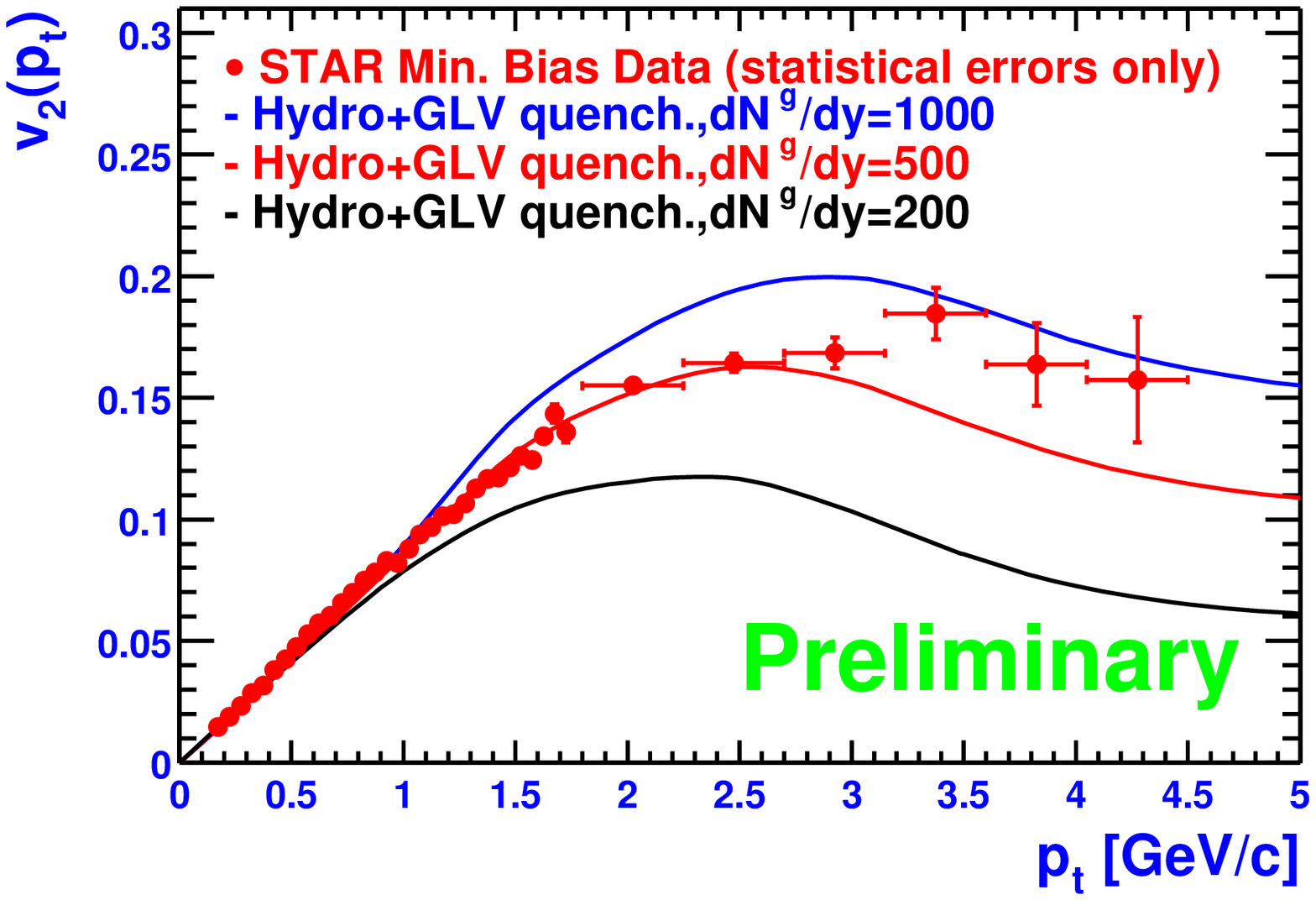,height=5.0cm,width=5.5cm}
\end{minipage}
\caption {Left panel: Elliptic flow vs beam energy ~\protect\cite{ceres-qm01}. 
          Right panel: Elliptic flow vs $p_T$ measured by STAR in Au-Au collisions at 
          $\sqrt{s_{NN}} = 130~GeV$ ~\protect\cite{harris-qm01}. Curves represent hydrodynamic 
          calculations with parton energy loss assuming different gluon densities 
          $dN^g/dy$~\protect\cite{gyulassy-vitev-wang}.}
\end{figure}
   
   Other interesting features appear in the $p_T$ dependence of $v_2$ shown in Fig. 3 (right panel)
\cite{harris-qm01}. Up to $p_T$ of $1~GeV/c$, hydrodynamic calculations, which
implicitly assume local thermal equilibrium, are in remarkably good agreement with the data,
contrary to the observations at the SPS where the experimental results are significantly below the 
hydrodynamic limit \cite{kolb}. 
At higher $p_T$, when pQCD starts to become significant,
$v_2$ appears to saturate. This behaviour can be reproduced by incorporating in the
hydrodynamic  model the energy loss of hard partons (jet quenching) in a dense medium with 
an initial gluon density of 
$dN^g/dy \sim 500 - 1000$ \cite{gyulassy-vitev-wang} \footnote{This might not be a unique
interpretation. In a very recent paper, the $v_2$ saturation at high $p_T$ is explained by the 
two-particle correlation of produced minijets \cite{kovchegov-tuchin}.}. This novel feature 
of jet quenching is further discussed in Section 6.
 
\vspace*{-0.3cm}        
\subsection{\underline {Particle interferometry}}
  Two-particle correlation with small relative momentum (HBT) is a useful tool to learn about the 
lifetime and size of the source at the time of particle emission. Since pions and kaons are emitted 
rather late, these correlations provide information about the space-time extent of the hadron system 
in the later stages of the collision.  
\begin{figure}[h!]
\vspace{-0.3cm} \centering
\epsfig{file=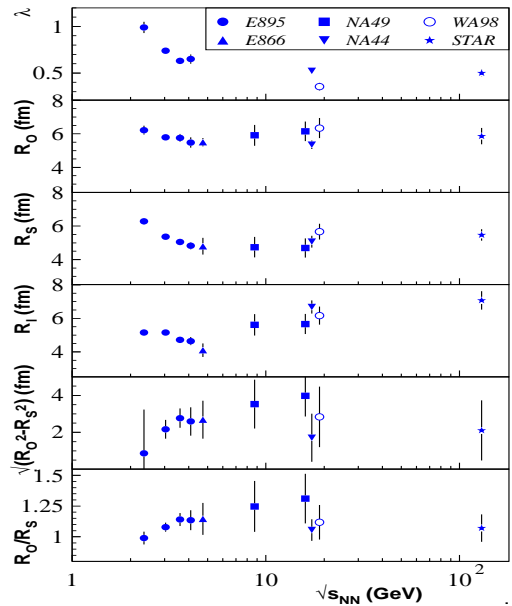, width=7.5cm,height=8.5cm}
\caption {Energy dependence of $\pi^-$ HBT parameters ~\protect\cite{star-hbt}.}  
\end{figure} 
The HBT technique has been widely used in ion experiments over the past decade at the AGS and 
SPS \cite{hbt-review} and first results from RHIC on pion interferometry are already available 
\cite{star-hbt,phenix-hbt}. The systematic behaviour of the source parameters --the radii 
$R_o, R_s, R_l$ and the coherence parameter $\lambda$--, 
as function of energy is displayed in Fig.4. The RHIC results are a big puzzle. All HBT parameters
have almost the same values as at the SPS. In particular, the larger source size $R_s$
and longer source lifetime reflecting itself in $R_o / R_s \geq 2$, expected if a QGP is formed 
\cite{rischke-gyulassy}, are not observed at RHIC. 
Instead, $R_s$ is very close to the SPS value and $R_o / R_s \sim 1$ again as at the SPS. These results are 
surprising and raise questions about the present understanding of the space-time evolution of relativistic
heavy-ion collisions.

\section {HADRONIC OBSERVABLES}
\vspace*{-0.3cm}
  Hadronic observables (spectra of identified particles, ratios of particle/antiparticle production, 
absolute particle yields) contribute, together with the global observables discussed in the previous 
section, to complete our view of the space-time evolution of RHI collisions. From their 
systematic studies, the picture has emerged of an initial high density state, which quickly reaches
equilibrium and cools down while undergoing collective expansion. Particles decouple when the
temperature and thus the particle density are low enough that interactions stop. We distinguish 
the chemical freeze-out when inelastic collisions cease to occur and particle abundances are frozen and 
the kinetic freeze-out, occuring later and at a lower temperature, when elastic processes stop and 
transverse momenta are frozen. 
  
\vspace*{-0.3cm}
\subsection{\underline {Chemical freeze-out}}
If we assume that an ideal hadron gas in thermal equilibrium is formed, then, in a statistical model
using the grand canonical ensemble, the final particle production
ratios are governed by only two independent variables, the temperature T and the baryochemical 
potential $\mu_B$ characterizing the system at chemical freeze-out \cite{pbm-sps}. This simple model 
works extremely well and reproduces all data available from SIS, AGS, SPS and also the new data from RHIC.
It is interesting to note that the enhanced production of strange and multi-strange hadrons, one of the 
first proposed signatures of QGP formation, is also described by the statistical model nearly as well as 
all other species. 

The parameters T and  $\mu_B$ derived from SPS  and RHIC data are listed in Table~1 \cite{pbm-sps,pbm-rhic}. 
\begin{table}[h]
\vspace*{-0.4cm} 
\caption{Chemical freeze-out parameters (temperature T and baryochemical potential  $\mu_B$) 
derived from SPS  ~\protect\cite{pbm-sps} and RHIC data  ~\protect\cite{pbm-rhic}}
\begin{center}
\begin{tabular}{lcc}
                       &     SPS        &   RHIC    \\ \hline                          
T ($MeV$)                &  168 $\pm$5    & 175 $\pm$7 \\
$\mu_B$ ($MeV$)          &  $\sim$270     &  51 $\pm$6 \\      
\end{tabular}
\end{center}
\end{table} 
 
\vspace*{-0.4cm} 
The temperatures are practically the same at RHIC and SPS, whereas 
$\mu_B$ is considerably lower at RHIC reflecting a lower net baryon density (see below). 
However, at both SPS and RHIC, the freeze-out parameters in the T vs  $\mu_B$ plane are close to 
the expected boundaries of the QGP phase transition. So presumably, prior to freeze-out, the system
was in a deconfined or at least in a mixed phase.

\vspace*{-0.3cm}
\subsection{\underline {Kinetic freeze-out}}  
  The conditions at kinetic freeze-out are derived from the transverse momentum spectra of 
identified particles. As an example, Fig. 5 shows the positive and negative $p, K, \pi$ spectra 
measured by PHENIX in central Au-Au collisions at $\sqrt{s_{NN}}=130~GeV$ \cite{phenix-ppg006}.
\begin{figure}[h!]
\vspace{-0.4cm}
\centering
\epsfig{file=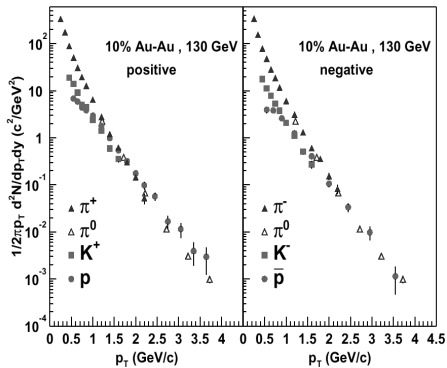,height=5.0cm,width=6.0cm}
\caption {Identified particle spectra measured by PHENIX in central Au-Au collisions at
           $\sqrt{s_{NN}}=130~GeV$ ~\protect\cite{phenix-ppg006}. }
\vspace{-0.5cm}
\end{figure}
The spectra exhibit an almost exponential shape with a temperature T (the inverse slope parameter) 
increasing with particle mass. This qualitative behaviour is independent of $\sqrt{s_{NN}}$. It has 
been observed all the way from BEVALAC up to RHIC energies 
and can be understood if one postulates that there is collective radial flow on top of the thermal 
chaotic contribution \cite{schnedermann93}, so that that the measured inverse slope 
$T = T_{fo} + m <\beta_T>^2 $. Particle spectra are then reproduced with only two parameters, 
characterizing the system at kinetic freeze-out, the freeze-out temperature  $T_{fo}$ and the 
collective radial flow velocity  $<\beta_T>$.  $T_{fo}$ is basically constant at $\sim 130~MeV$ 
from AGS to RHIC as expected if particles decouple at a given density. A constant radial flow velocity
of  $<\beta_T> \sim 0.4$ has been established at AGS and SPS energies, whereas the first RHIC results 
point to a higher value of $\sim 0.6$ indicating higher initial pressure in 
the system. The resulting larger inverse slopes lead to an interesting feature not observed
at lower energies. The $\overline{p}$ and $\pi^-$ (as well as the $p$ and $\pi^+$) yields become 
comparable at $p_T \approx 2 GeV/c$. 

At the SPS, the $\Omega$ and $J/\psi$ (and probably also the $\Xi$) particles follow 
a different pattern \cite{star-p-qm01}.
They exhibit stronger slopes than expected from the previously quoted formula. This is taken as evidence
that these particles decouple earlier from the system due to their relatively large mean free 
path. It will be very interesting to see whether similar data from RHIC will follow 
the same pattern. 

\vspace*{-0.3cm}
\subsection{\underline {Baryon yields}} 
    Baryon yields have consistently attracted interest: (i) the net baryon yield at mid-rapidity 
is an important observable allowing to study the amount of baryon stopping (the transport of 
participating nucleons to mid-rapidity) that occurs early in the 
collision. (ii) the mechanism of baryon stopping is a an unsolved question and a topic of recent debate 
\cite{dima-gluon-junction,vance-gluon-junction}. (iii) the total baryon density plays an important 
role  in connection with in-medium modification of light vector mesons and low-mass dilepton production 
as emphasized recently \cite{rapp-total-baryon} (see Section 4).
 
The $\overline{p}/p$ and $\overline{\Lambda}/\Lambda$ ratios increase dramatically from values of
$\sim0.1$ at SPS to values of $\sim 0.7$ at RHIC
\cite{phenix-ppg006,star-lambda,star-p-qm01,phenix-ppg012}. Clearly the ratios do not yet reach 
unity. The system approaches net baryon free conditions, but remarkably, still $\sim5\%$ 
of the participating nucleons are transported over 5 units of rapidity as shown in Table~2.

From the inclusive (i.e. non-corrected for feed-down) measurement of $p$ and  $\overline{p}$ one can 
get a crude estimate of the total baryon density (under the assumption that $p$ and $n$ have the same 
stopping probability, that  $p\overline{p}$ and $n\overline{n}$ pairs are produced in equal quantities 
and that higher mass baryons (mainly lambdas and sigmas) finally end up as $p,\overline{p},n$ 
or $\overline{n}$). The relevant quantities for RHIC are listed in Table 2. A similar estimate at
\begin{table}[h]
\caption{Net and total baryon density at RHIC }
\hskip4pc\vbox{\columnwidth=26pc
\begin{tabular}{lc}
                                                      &   RHIC Au-Au at $\sqrt{s_{NN}}=130GeV$  \\ \hline
$dN(p)/dy$                                            &  28.7$^{a}$  \\
$dN({\overline p})/dy$                                &  21.1$^{a}$         \\
Net protons $dN(p - {\overline p})/dy$                &    8.6          \\
Participating Nucleons ($p - {\overline p})A/Z$       &   21.4         \\  \hline
Produced Baryons ($p, {\overline p},n, {\overline n}$)&   80.4             \\ \hline
Total Baryon Density $dN_B/dy$                        &   102           \\ 
\end{tabular}
}
\vspace*{0.2cm} 
a) inclusive $p$ and  $\overline{p}$ rapidity density from \cite{phenix-ppg006}
\end{table}
the maximum SPS energy of  $\sqrt{s_{NN}}=17~GeV$, derived from the feed-down corrected 
$p$, $\overline{p}$ yields and the  $\Lambda$, $\overline{\Lambda}$ yields \cite{na49-lambdas} gives a 
total baryon density of $dN_B/dy = 110$. It appears therefore that, contrary to previous expectations, 
the total baryon density is 
approximately the same at SPS and RHIC, the strong decrease in nuclear stopping being compensated by copious
production of baryon-antibaryon pairs. 

\vspace*{-0.2cm} 
\section{LOW-MASS $e^+e^-$ PAIRS and PHOTONS}
   Dileptons and photons are unique probes to study
the dynamics of relativistic heavy-ion collisions \cite{shuryak}.
 The interest stems from their relatively large mean free path. As a consequence, they
can leave the interaction region without final state interaction, carrying information about 
the conditions and properties of the matter at the time of their production
and in particular at the early stages 
when deconfinement and chiral symmetry restoration 
have the best chances to occur. 

 The prominent topic of interest, both in dileptons and photons, 
is the identification of thermal radiation emitted from 
the collision system. This radiation should tell us the nature of the matter formed, 
a quark-gluon plasma (QGP) or a high-density hadron gas (HG). 
The physics potential of low-mass dileptons is further augmented by their
sensitivity to chiral symmetry restoration. The $\rho$-meson is the prime agent here.
Due to its very short lifetime  ($\tau~=~1.3~fm/c$) compared to the typical fireball 
lifetime of $\sim10~fm/c$ at SPS energies, most of the $\rho$ mesons 
decay inside the interaction region providing a unique 
opportunity to observe in-medium modifications of particle properties (mass and/or width)
which might be linked to chiral symmetry restoration. 
The situation is  different for the $\omega$ and $\phi$ mesons. Because 
of their much longer lifetimes they predominantly decay outside the interaction region after 
having regained their vacuum properties.
The  $\omega$ and $\phi$ mesons remain nevertheless important messengers: the 
undisturbed   $\omega$ can serve as a reference and the  $\phi$ with its $s{\overline s}$
content is a probe of strangeness production.

  For the moment, the stage for low-mass dileptons belongs to CERN and more specifically to the 
CERES experiment.
CERES has systematically studied the production of $e^+e^-$ pairs in the mass 
region $m \leq 1.0~GeV/c^2$, with measurements of p-Be,
p-Au, \cite{pbe-ee,pbe-eegamma}, S-Au \cite{prl95} and Pb-Au \cite{pbau95-plb,bl-qm99,ceres-qm01}.
Whereas the $p$ data are well reproduced by the known hadronic sources, a strong enhancement 
is observed in the mass region $ 200 < m < 600~MeV/c^2$
both in the S and Pb data with respect to those sources scaled to the nuclear case with
the event multiplicity. Fig. 6 (left panel) shows the results from Pb-Au collisions 
at $40~A~GeV$ with an enhancement factor of $5.1\pm1.3(stat.)$ relative to the hadronic cocktail. 
\begin{figure}[h!]
\vspace*{-0.3cm}  
\begin{minipage}[h!]{55mm}
\hspace*{1.0cm}
\epsfig{file=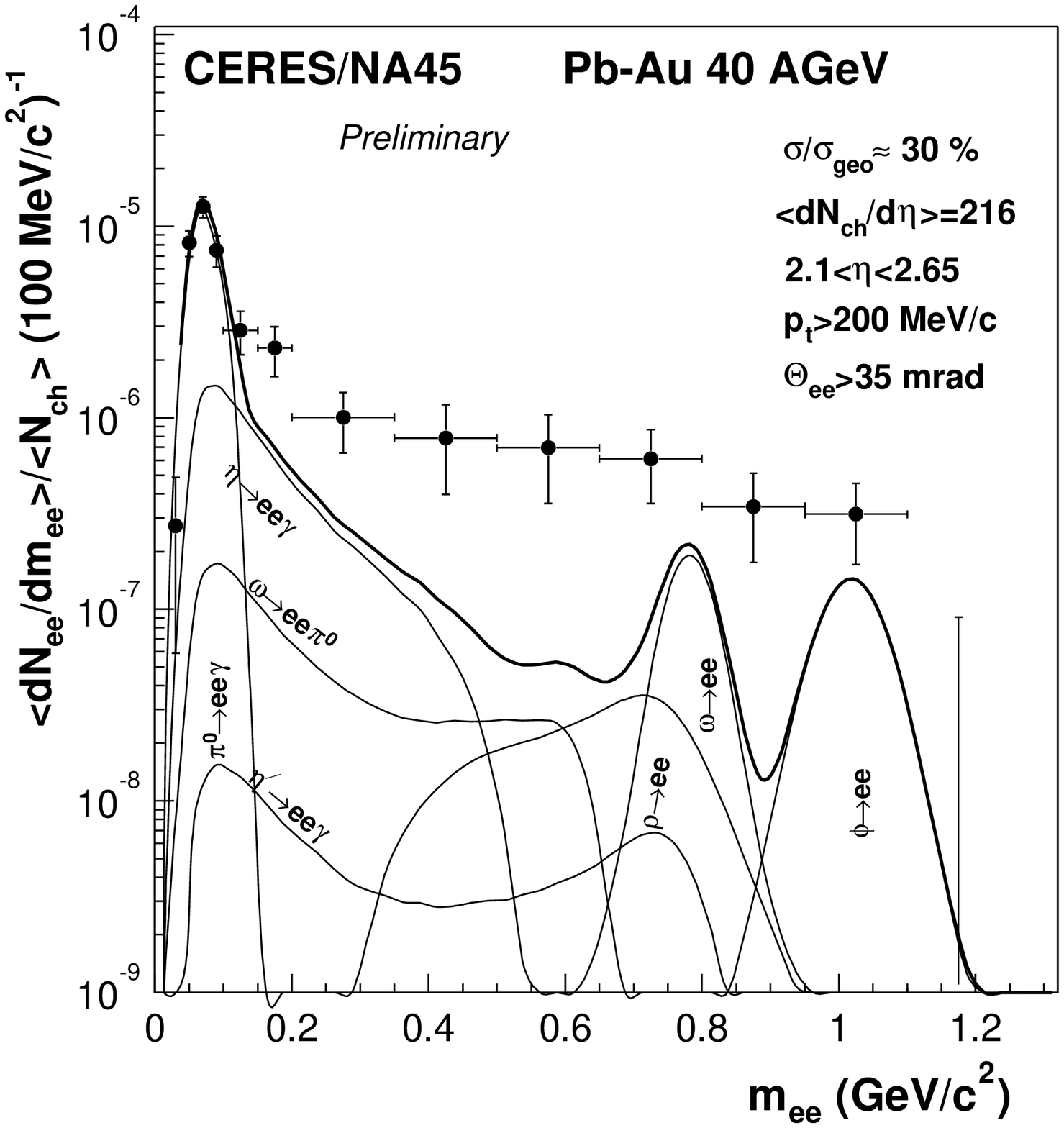,height=5.5cm}
\vspace{-0.4cm}
\end{minipage}
\begin{minipage}[h!]{55mm}
\vspace{-5.9cm}
\hspace*{3.0cm}
\begin{rotate}{-90}
\epsfig{file=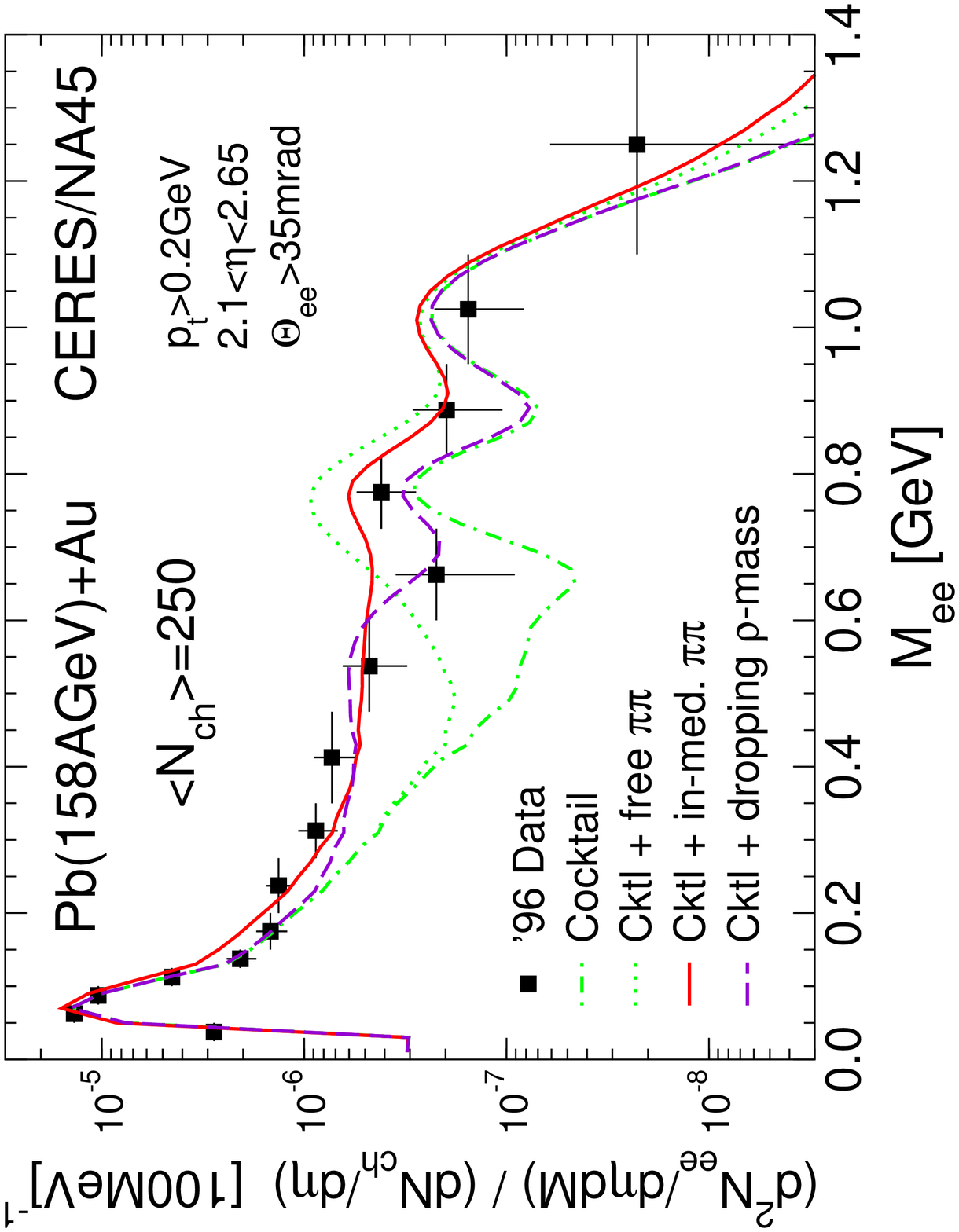,width=5.9cm,height=6.0cm}
\end{rotate}
\end{minipage}
\caption {Left panel: Invariant mass $e^+e^-$ spectrum measured by CERES in $40~A~GeV$ Pb-Au collisions
          ~\protect\cite{ceres-qm01}. The figure also shows the summed (thick solid 
          line) and individual contributions from hadronic sources.
          Right panel: Comparison of CERES 96 results from Pb-Au collisions at $158~A~GeV$ with 
          various theoretical approaches (see text).}
\vspace*{-0.4cm}
\end{figure}
Within uncertainties, this is consistent with, or even larger than,
the enhancement factor of
$2.9\pm0.3(stat.)\pm0.6(syst.)$ obtained from the combined 95-96 Pb-Au data at $158~A~GeV$.
Further studies on the latter demonstrate that the enhancement is more pronounced at low pair $p_T$
and increases faster than linearly with the event multiplicity. In all cases, with the S and Pb beam, 
the enhancement sets at $m~\geq~2m_{\pi}$. 

  The enhancement of low-mass dileptons has triggered a wealth of theoretical activity 
(for a comprehensive review see \cite{rapp-wambach}).  
There is consensus that  a simple superposition of $pp$ collisions cannot explain the data and that 
an additional source is needed. The pion annihilation channel 
($\pi^+\pi^- \rightarrow \rho \rightarrow l^+l^-$), 
obviously irrelevant in $pp$ collisions, has to be added in the nuclear case. This channel
accounts for a large fraction of the observed enhancement (see line cocktail + free $\pi\pi$ in Fig. 6 
right panel) and provides first evidence of thermal radiation from a dense hadron gas. However, in order to 
quantitatively reproduce the data in the mass region $0.2 < m_{e^+e^-} < 0.6~GeV/c^2$, it was 
found necessary to include in-medium modifications of the $\rho$ meson. Li, Ko and Brown \cite{li-ko-brown}, 
following the original Brown-Rho scaling \cite{brown-rho}, proposed a decrease of the $\rho$-mass in 
the high baryon density of the  fireball, as a precursor of chiral symmetry restoration, and 
achieved excellent agreement with the CERES data (see the line cocktail + dropping mass in Fig. 6 
right panel). 
 
 Another avenue, based on effective Lagrangians, uses the broadening of the $\rho$-meson spectral function 
resulting from its propagation in the medium, mainly from scattering off baryons 
\cite{wambach}, and achieves also an excellent reproduction of the CERES data (see line cocktail + 
in-medium $\pi\pi$ in Fig. 6 right panel). 
The success of these two different approaches, one relying on quark degrees of freedom and the other on a 
pure hadronic model, has attracted much interest. Rapp raised, and provided support for,
the hypothesis of quark-hadron duality down to low-masses by showing that in a high density state 
the dilepton 
production rates calculated with the in-medium  $\rho$-meson spectral function are very similar to the 
$q{\overline q}$ annihilation rates calculated in pQCD \cite{rapp-duality-qm99}.
It is interesting to note that the $40~A~GeV$ data is also equally well reproduced by the 
two approaches \cite{ceres-qm01}. The accuracy of the data
does not allow to discriminate among the two. The CERES run of year 2000 taken 
with improved mass resolution and good statistics may allow 
to do that. 
 
  The extension of these studies to RHIC energies should be very interesting. The total baryon density,
which is the key factor responsible for in-medium modifications of the $\rho$ meson both in the 
dropping mass and the collision broadening scenarios, is almost as high at RHIC as at SPS (cf.
Section 3.3), contrary to previous expectations.  Updated RHIC predictions that incorporate this and
other acquired knowledge on global and hadronic observables, show indeed that the enhancement of low-mass 
$e^+e^-$ pairs persist at the collider with at least comparable strength \cite{rapp-bahamas}. The 
calculations further predict in-medium modifications of the $\omega$ and $\phi$ mesons. These are
much less dramatic than in the case of the $\rho$ meson but should nevertheless 
be readily observable with the excellent mass resolution of the PHENIX detector. First results on 
single electrons from PHENIX are already available \cite{phenix-ppg011} and results  
on the $\phi$ meson from the second RHIC run are expected soon. The combined analysis of 
$\phi \rightarrow K^+K^-$ and $\phi \rightarrow e^+e^-$ within the same apparatus should provide 
a very powerful diagnostic tool to make evident in-medium effects.

  Direct photons are expected  to provide analogous information to thermal dileptons. However, 
the physics background for real photons is larger by orders of magnitude compared to dileptons  
making the measurement of photons much tougher and less sensitive to a new source. And indeed,  
only a small direct photon signal of $\sim 20\%$ has been observed in central Pb-Au collisions by
the WA98 experiment \cite{wa98-photons}. The better conditions at RHIC, larger particle densities
and hence larger initial temperatures should provide a more compelling evidence of this key
signal.

\vspace*{-0.2cm} 
\section{$J/\psi$ SUPPRESSION}
  The melting of charmonium states in a deconfined state of matter is one of the
earliest signatures of QGP formation \cite{satz-matsui} and has provided one of the most
exciting sagas of the SPS programme. Already in the first runs with O and S beams, the
NA38/NA50 experiment observed a suppression of $J/\psi$ which immediately captured the interest. 
Intensive theoretical efforts were devoted to explain the effect within 
conventional physics 
\begin{wrapfigure}[16]{r}{5cm}
\epsfxsize=12pc
\vspace{-0.6cm}
\epsfbox{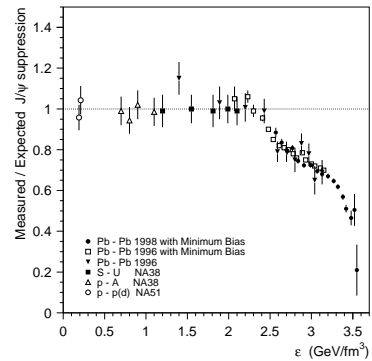}
\vspace{-0.6cm}           
\caption{Measured over expected 
       (assuming normal absorption cross 
        section) $J/\psi$ yield vs. energy 
         density ~\protect\cite{na50-jpsi}.}
\end{wrapfigure}
 and it soon appeared that all experimental data, including systematic 
studies of $pp$, $pA$ and $SU$ collisions, can be reproduced by invoking the absorption in nuclear
medium of the $c\overline{c}$ pair before it forms a $J/\psi$, with a cross 
section $\sigma_{abs} = 6.4 \pm 0.8 mb$~\cite{na38-jpsi}. 

   However, a different behavior is observed with the $158~A~GeV$ Pb beam. Whereas peripheral 
collisions seem to  follow the regular absorption pattern, an anomalously larger suppression occurs 
at central collisions characterized by impact parameters $b < 8~fm$ or $E_T~>~40~GeV$ which 
can be translated into a Bjorken energy density $\epsilon > 2.3~GeV/fm^3$ (see Fig. 7) \cite{na50-jpsi}.
The data in this figure exhibit a two-step suppression pattern which NA50 attributes to the successive 
melting of the $\chi_c$ and directly produced $J/\psi$ mesons in a quark-gluon plasma scenario. 
  
The same data normalized to the Drell-Yan cross section are shown in Fig.~8 
as function of $E_T$. The two-step pattern can be discerned here at $E_T$ values of
$\sim30$ and $\sim100~GeV$. Most published calculations, based on conventional hadronic models 
including the effect of absorption by comovers, fail to reproduce the results as illustrated in 
the left panel of Fig.~8. The recent calculations of Capella et al. \cite{capella} are in much better 
agreement with the data and fail only at the most central collisions (middle panel).
\begin{figure}[h!]
\vspace{-0.7cm}
\includegraphics*[width=4.2cm]{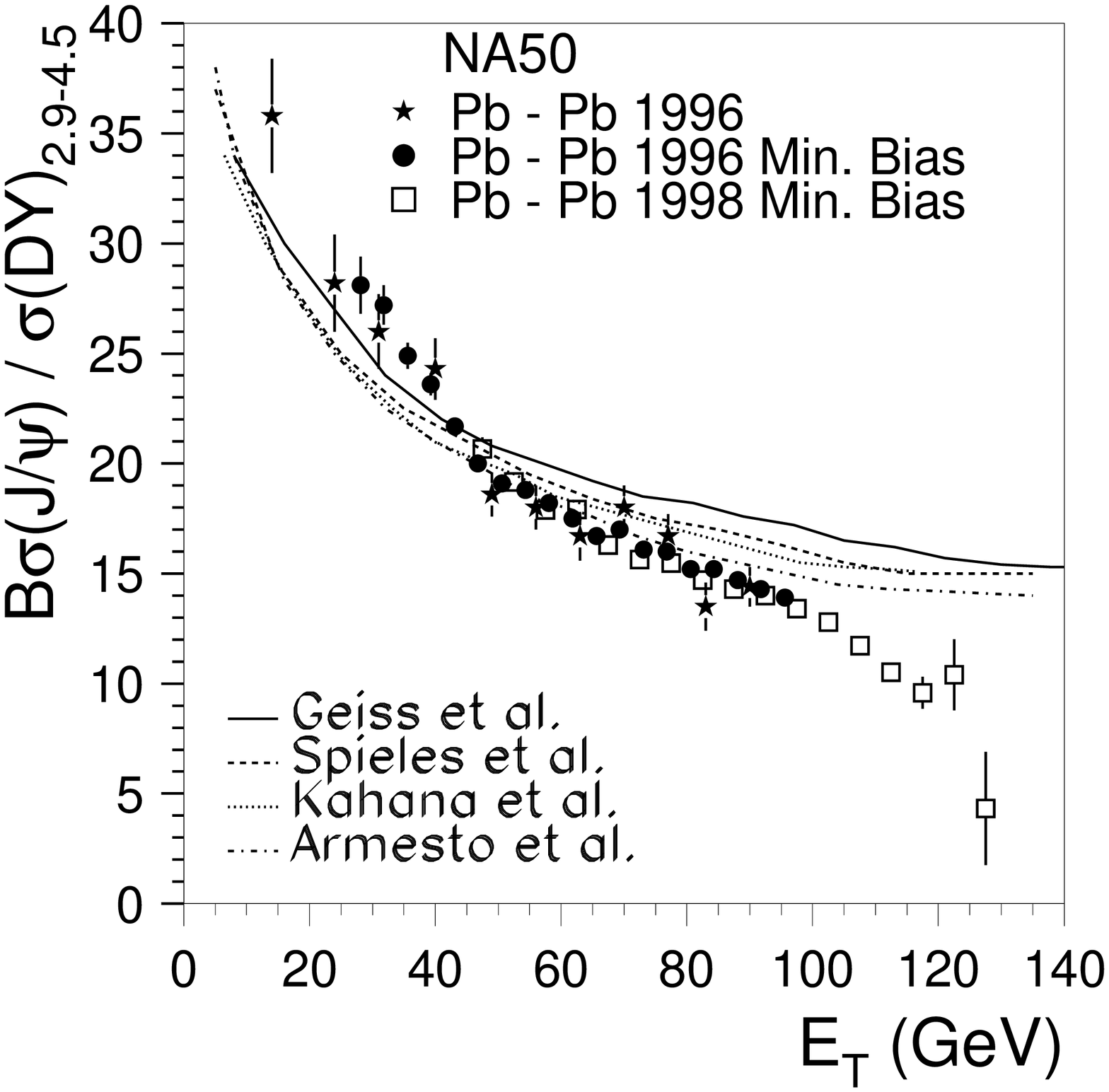}
\hspace{1.0cm}
\includegraphics*[width=4.2cm]{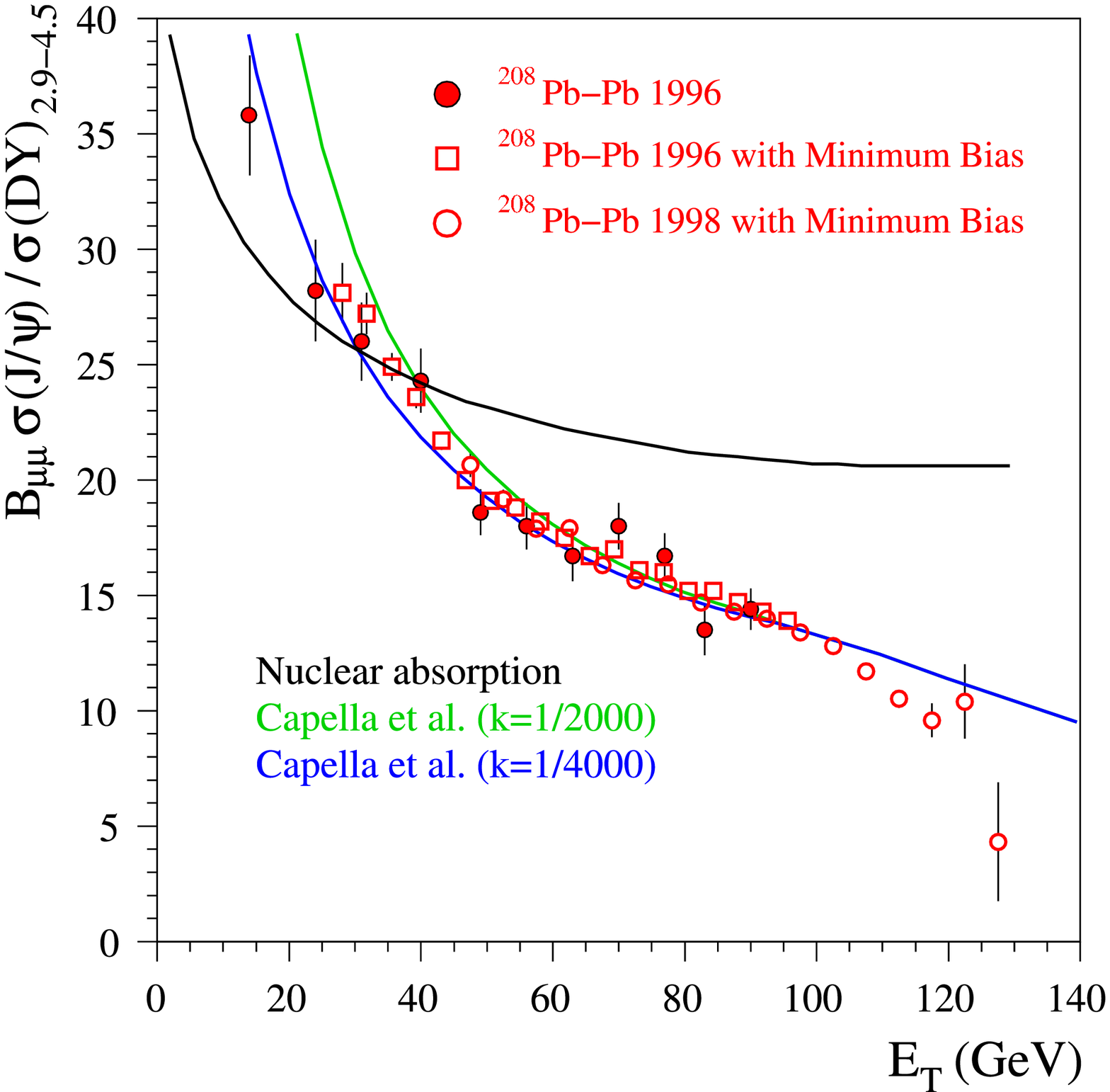}
\hspace{1.0cm}
\includegraphics*[width=4.2cm]{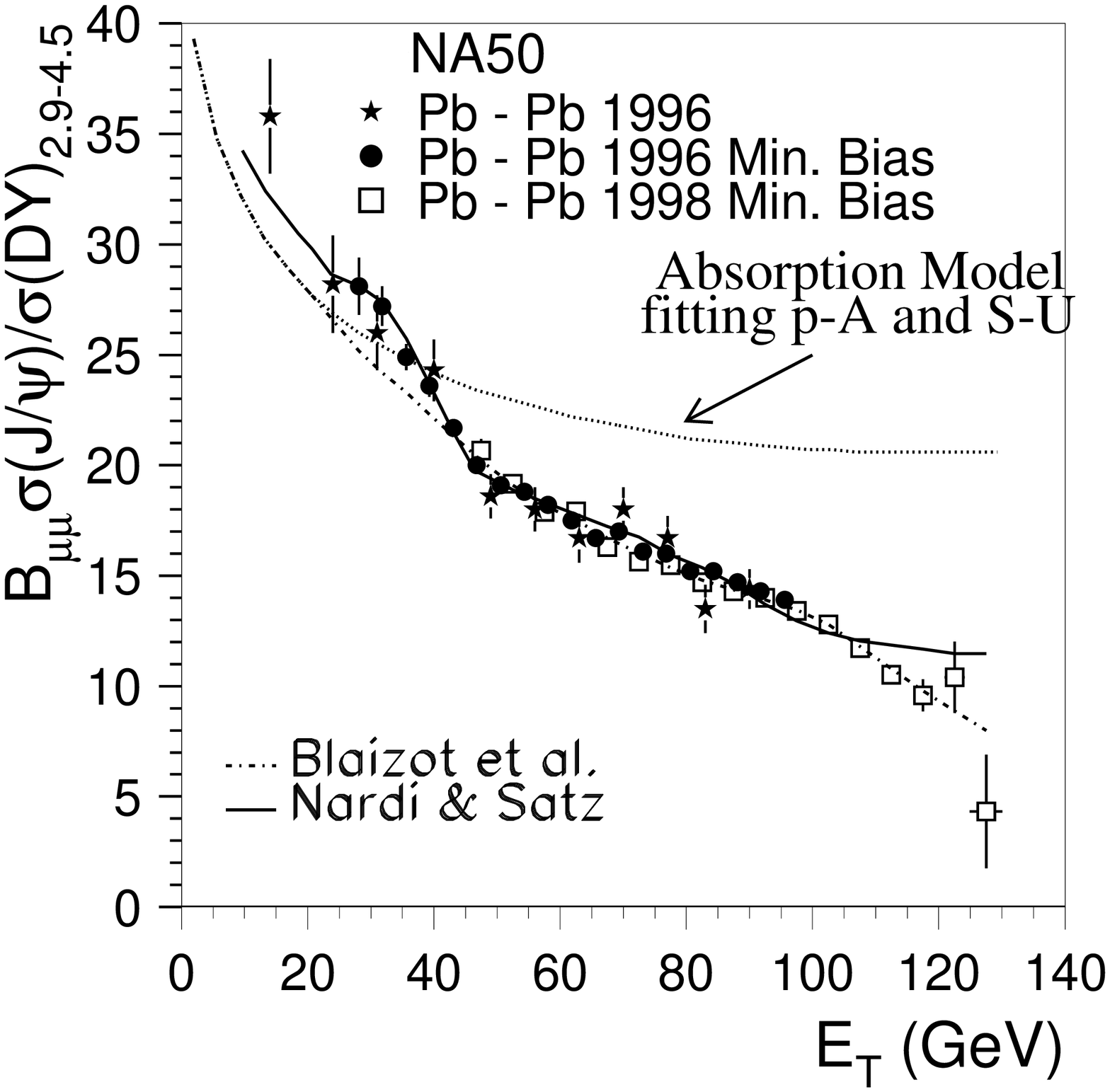}
\caption{$J/\psi$ over Drell-Yan cross section vs. $E_T$ measured by NA50 in Pb-Pb
         collisions at  $\sqrt{s_{NN}}~=~17.2~GeV$ in comparison to various theoretical approaches 
         using conventional physics (left and middle panels) and quark matter
          formation (right panel) (taken from ~\protect\cite{bordalo-qm01}).}
\end{figure}
On the other hand, quite good agreement over the whole $E_T$ range is achieved by models assuming 
QGP formation (right panel) \cite{blaizot2000,nardi-satz}. The model of Blaizot et al. reproduce the 
data remarkably well by invoking  $J/\psi$ suppression whenever the energy density exceeds the critical 
value for deconfinement (first step)  together with fluctuations of the transverse energy for the most 
central collisions (second step) \cite{blaizot2000}. 

    The extrapolation of these results to RHIC energies is not at all clear and ranges from total
suppression up to enhancement ! \cite{jpsi-enhancement} thus ensuring another exciting chapter 
on the $J/\psi$ saga. First glimpse on $J/\psi$ production at RHIC is expected from the PHENIX results 
of year 2 run. 

\section{SUPPRESSION OF LARGE $p_T$ HADRONS}
 With the energies available at RHIC, one order of magnitude higher than at the SPS,
new channels and probes become accessible for matter diagnostics.
In particular, energy loss through gluon radiation of high $p_T$ partons resulting from initial hard
scattering and jet production has been predicted as a possible signature of deconfined matter.
The phenomenon, commonly referred to as jet quenching, has been extensively studied over the last 
decade \cite{wang-miklos-quenching} and should manifest itself as a suppression of high  $p_T$ particles. 
Such a suppression has been observed already in the first low-luminosity RHIC run by the
two large experiments STAR \cite{harris-qm01} and PHENIX \cite{phenix-ppg003} and certainly constitutes 
the most interesting result from RHIC so far. The suppression is evidenced by plotting the so-called 
nuclear modification factor\cite{wang-wang} defined as the ratio of $AA$ to $pp$ $p_T$ spectra, 
scaled by the number of binary collisions:   

\[ R_{AA}(p_T) = \frac{d^2\sigma_{AA}/dp_T d\eta} { < N_{bin} > d^2\sigma_{pp}/ dp_T d\eta}  \]  

In the absence of any new physics this ratio should be equal to 1 at the high $p_T$
characteristic of hard processes. At low $p_T$,  dominated by soft processes which scale with the number of 
participants, the ratio is expected to be lower, e.g. for central collisions  
$R_{AA}(0) \approx 0.5 N_{part} / N_{bin} \approx 0.2$. The STAR and PHENIX results for central
Au-Au collisions at $\sqrt{s_{NN}} = 130~GeV$ are shown in
Fig. 9. In both cases the $pp$ data at $130~GeV$ were derived from interpolation of $pp$ data
\cite{ua1,isr,cdf} at lower and higher energies. 
\begin{figure}[h]
\vspace{-0.2cm}
\begin{minipage}[t]{55mm}
\hspace*{1.0cm}
\epsfig{file=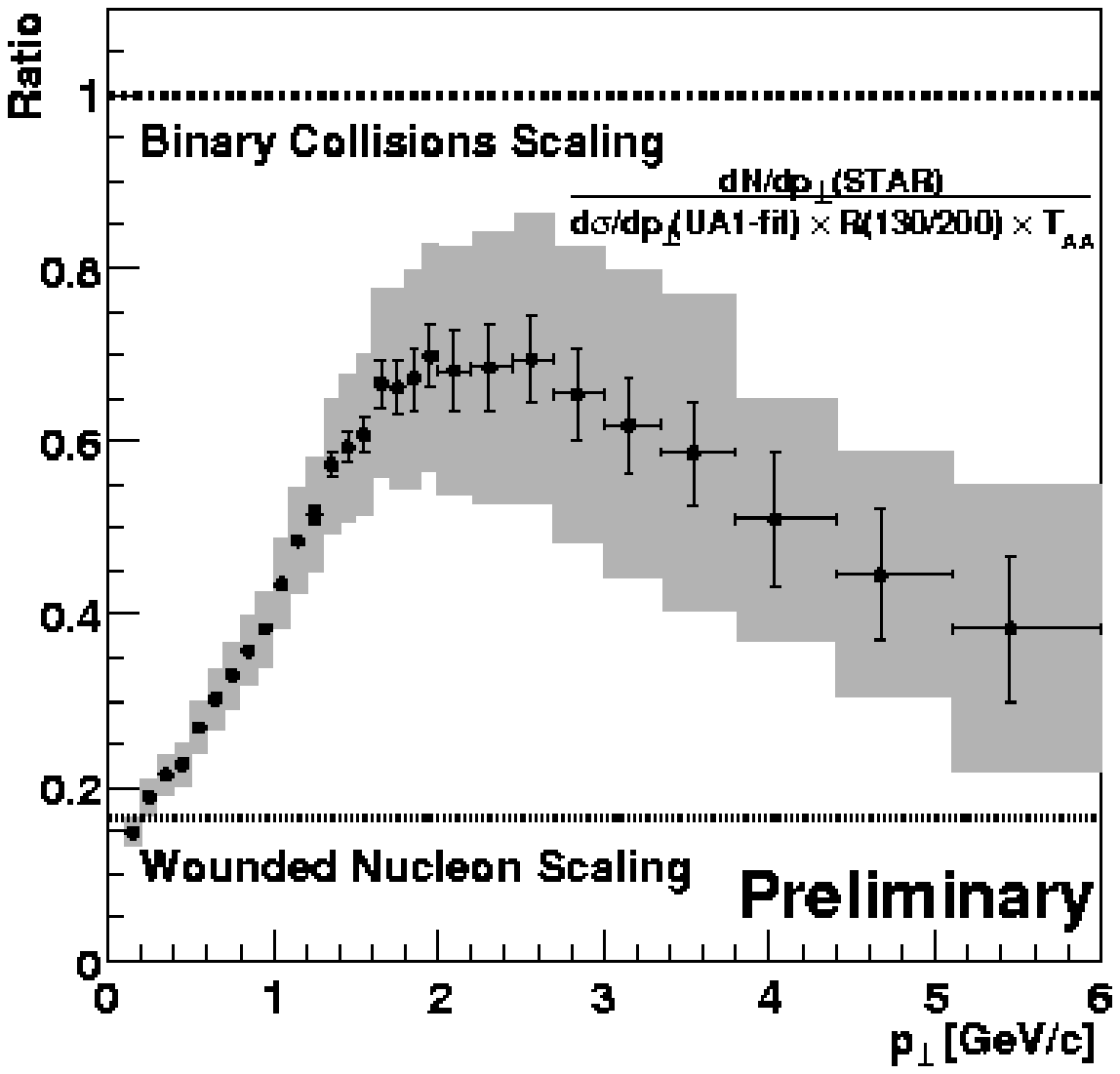,height=5.3cm}
\vspace{-0.6cm}
\end{minipage}
\begin{minipage}[ht]{55mm}
\vspace{-5.0cm} 
\hspace*{3.0cm}
\epsfig{file=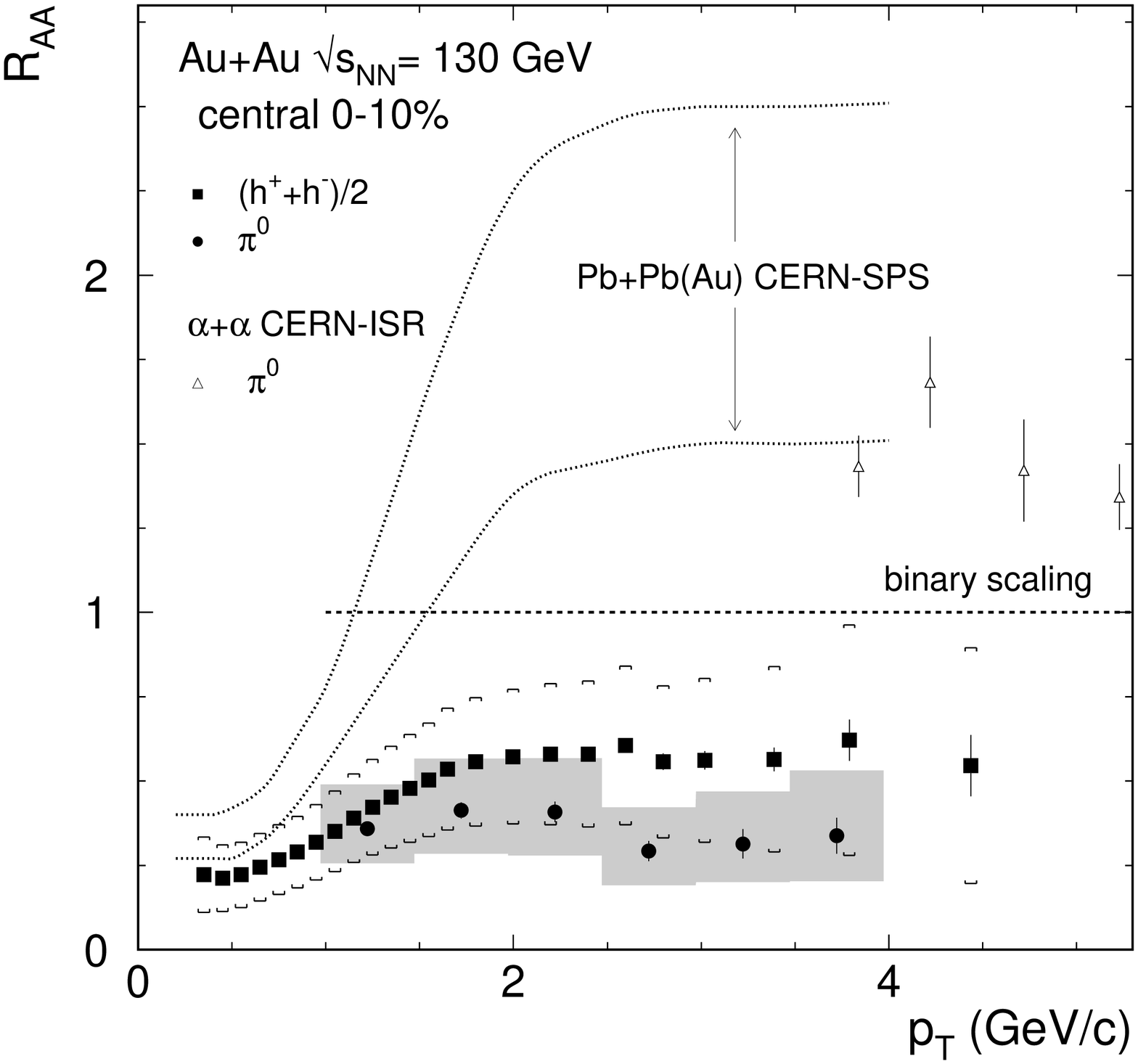,height=5.7cm}
\end{minipage}
\vspace{-0.2cm}
\caption {High $p_T$ suppression in Au-Au collisions at  $\sqrt{s_{NN}} = 130 GeV$ 
          measured by STAR for negative hadrons (left panel) ~\protect\cite{harris-qm01} and in PHENIX 
          for negative hadrons and $\pi^o$ (right panel) ~\protect\cite{phenix-ppg003}.}  
\end{figure}
 STAR shows the ratio for negative hadrons and PHENIX for negative hadrons as well as identified 
$\pi^o$. As expected, for low $p_T$ the ratio is small, $\approx 0.2$. It raises with  $p_T$, getting 
close to unity and then falls off at higher $p_T$ in the STAR data whereas it seems to flatten
in the PHENIX data. The suppression appears to be stronger for
$\pi^o$ than for charged particles. This implies that the ratio $\pi^o/h$ is considerably 
smaller than unity and reflects the large $p,\overline{p}$ contribution to the charged 
particle spectra at high momentum (cf. Section 3.2).
A similar suppression pattern appears when the ratio of central to peripheral collisions, each divided 
by its corresponding value of $N_{bin}$, is plotted as function of $p_T$ \cite{phenix-ppg003}.
This behaviour is in marked contrast to the observations in Pb-Pb and Pb-Au collisions 
at the SPS (the solid lines in Fig. 9 right panel represent the band of uncertainty of the CERN results)
where the ratio overshoots unity and saturates at high  $p_T$, like in pA collisions \cite{wang-wang} due 
to the well known Cronin effect \cite{cronin}. The high $p_T$ suppression observed at RHIC can 
be quantitatively reproduced in terms of jet quenching if an average energy loss of $dE/dx 
\approx 0.25~GeV/fm$ is assumed within a parton model\cite{wang-qm01}.

   In the context of the new RHIC data, jet quenching has been discussed  in three different topics 
(cf. Sections 2.1, 2.2 and this Section). Whereas the phenomenon helps describing the observed effects
in elliptic flow and high  $p_T$ suppression, the expected increase in particle production has not
been observed (cf. Section 2.1). Fine tuning of the theoretical treatment to achieve consistent 
description of all these observables  together with new RHIC data allowing to reach much higher 
$p_T$ values and reference data on $pp$ and $pA$ measured within the same apparatus will be very 
valuable to consolidate these intriguing results.

\vspace{-0.3cm}
\section{CONCLUSION}
   In a relatively short run, the RHIC experiments have produced an impressive amount of
results and much more is expected over the next years. The second RHIC run has just been 
completed with a recorded luminosity almost two orders of magnitude larger than in year-1.
This, together with the still ongoing yield of interesting results from the SPS programme, 
places the field of relativistic heavy-ion collisions at a very unique phase with exciting 
physics prospects.

\end{document}